\documentclass[twocolumn,prb,superscriptaddress,showpacs]{revtex4}
\usepackage{epsfig}
\usepackage{times}
\usepackage{amsmath}

\begin{document}

\title{Spin polarization of the $\nu=5/2$ quantum Hall state}

\author{A.~E.~Feiguin}
\affiliation{Microsoft Station Q, University of California, Santa Barbara, California 93106, USA}
\affiliation{Condensed Matter Theory Center, Department of Physics, University of Maryland, College Park, Maryland 20742, USA}
\author{E.~Rezayi}
\affiliation{Department of Physics, California State University, Los Angeles, California 90032, USA }
\author{Kun~Yang}
\affiliation{NHMFL, and Department of Physics, Florida State University, Tallahassee, Florida 32306, USA }
\author{C.~Nayak}
\affiliation{Microsoft Station Q, University of California, Santa Barbara, California 93106, USA}
\author{S.~Das Sarma}
\affiliation{Condensed Matter Theory Center, Department of Physics, University of Maryland, College Park, Maryland 20742, USA}

\date{\today}

\begin{abstract}
We numerically study the spin polarization  
of the fractional 
quantum Hall state at filling factor $\nu=5/2$. By using both exact 
diagonalization and the Density Matrix Renormalization 
Group (DMRG) methods on the sphere, we are able to analyze more values of
partial spin polarization (in addition to fully-polarized and unpolarized) than
any previous studies.
We find that for the Coulomb interaction the exact finite-system 
ground-state is fully polarized, 
for shifts corresponding to both the 
Moore-Read Pfaffian state and its particle-hole conjugate (anti-Pfaffian). This result is 
found to be robust against small variations of the 
interaction and change of shift. The low-energy excitation spectrum is consistent with spin-wave excitations of 
a fully-magnetized ferromagnet.
\end{abstract}

\maketitle

\section{Introduction}

The most striking feature of the Laughlin state describing the fractional quantum
Hall (FQH) effect at filling fraction $\nu=1/3$ \cite{Laughlin83} is the
appearance of quasi-particle excitations with fractional charge and fractional
statistics. The idea of particles that do not behave as fermions or bosons,
something that can occur in two spatial dimensions, is still a reason for
wonder, and a motivation for seeking phases of matter with exotic excitations
in low dimensions. The Laughlin wavefunction served as a foundation to explain all
the odd-denominator incompressible
FQH states \cite{Haldane83,Halperin84,Jain89,Read90,Wen92}.
However, it does not include the possibility of an even-denominator state.
Therefore, the quantum Hall plateau observed at $\nu=5/2$
\cite{Willett87,Pan99b,Eisenstein02,Xia04,Pan08,Choi08} poses a special
challenge.

While various theories have been proposed for this state
\cite{Haldane-Rezayi,331,Moore-Read,Greiter92,Park,Levin07,Lee07,Gunnar08,Yang08},
much of the excitement has been generated by the possibility
that it is a non-Abelian topological state.
In ground-breaking work, Moore and Read \cite{Moore-Read} proposed 
the Pfaffian wavefunction as a description of electrons
in an incompressible half-filled Landau level. Greiter {\it et al.} \cite{Greiter92}
conjectured that this ground-state may be realized at $\nu=5/2$.
Recently, it was noted that there is another possible state, 
the so-called anti-Pfaffian \cite{Lee07,Levin07}, which
would be degenerate in energy with the Pfaffian
state in the absence of Landau level mixing.
Since excitations above both the Pfaffian \cite{Nayak96,Read96,Read00,Ivanov01}
and anti-Pfaffian \cite{Lee07,Levin07,Peterson08b} ground-states
are non-Abelian anyons, it has been suggested \cite{DasSarma05}
that the $\nu=5/2$ plateau can be a platform for topological quantum
computation. Therefore, determining the nature of the $\nu=5/2$ state
has gained additional urgency, beyond FQH physics \cite{Nayak08}.

In order to set a context for the importance of our theoretical numerical
study of the 5/2 spin polarization question, we first briefly describe the
highly confusing experimental status of the subject.  Immediately
following the original discovery of the 5/2 FQHE, Eisenstein et al. \cite{Eisenstein88}
found that the application of a modest in-plane magnetic field destroys
the FQHE.  This was interpreted quite naturally as direct evidence for the
5/2 FQH state being spin-unpolarized, leading to proposed spin-singlet
wavefunctions \cite{Haldane-Rezayi} describing the 5/2 FQH state which, however, turned out
to have very poor overlap with the exact numerical wavefunction.  All
subsequent measurements \cite{Lilly99,Pan99a} of the 5/2 FQHE in the presence of an
in-plane magnetic field have verified its suppression in the presence of
even a weak in-plane magnetic fields.  The most direct interpretation of
such an in-plane field induced destruction of the 5/2 FQHE as arising from
the Zeeman splitting induced spin-polarization effect (i.e. the original
unpolarized FQH state becoming spin-polarized under the in-plane field)
becomes questionable, however, when one realizes that experimentally the
5/2 FQHE is observed over a very large range of perpendicular magnetic
fields, ranging from 2T \cite{Dean08} to 12T \cite{Pan00},  and therefore, the 5/2 FQHE can
obviously survive very large spin-polarizations!  A more plausible
scenario is that the in-plane magnetic field induced destruction of the
5/2 spin polarization arises \cite{Morf98, Haldane-Rezayi} from the orbital
coupling \cite{Peterson08} of the in-plane field and not at all from the Zeeman
coupling which depends on the total magnetic field.  Efforts \cite{Jim} to
directly measure the 5/2 spin-polarization through the resistive NMR
technique have so far been unsuccessful although similar measurements \cite{Tracy07,Li09}
at $\nu=1/2$ in the lowest Landau level have unambiguously established the
spin-unpolarized (or partially-polarized) nature of the (non-FQH) 1/2 state
in weak magnetic fields (up to 5-8T, much higher than magnetic fields
where the 5/2 FQHE is routinely observed).  Taken together, all of this
experimental evidence provides a highly conflicting picture for the
spin-polarization of the 5/2 FQH state, with both spin-polarized and
spin-unpolarized (certainly partially-polarized) states being plausible,
particularly at low magnetic fields.  

The existence of non-Abelian quasiparticles at $\nu=5/2$
depends on (at least) the following premises:
(i) Coulomb repulsion in the second LL (SLL) has a form conducive
to pairing and (ii) the electrons are fully spin-polarized.
There is strong evidence from numerics that (i) is satisfied
\cite{Morf98,Rezayi00,Feiguin,Xin06,Gunnar08,Peterson08b,Storni08} (especially when finite
layer thickness is taken into account \cite{Peterson08}).
Recent experiments which are consistent with a quasiparticle
charge $e/4$ \cite{Dolev08,Radu08} give further support to this
hypothesis, but cannot rule out Abelian paired states which also could have
$e/4$ quasiparticle charge. However, there is less evidence that (ii) holds.
In GaAs, the Zeeman energy is approximately 50 times smaller than
the cyclotron energy as a result of effective mass and $g$-factor
renormalizations, so the magnetic field need not fully polarize
the electron spins. Electron-electron interactions, which are roughly
comparable to the cyclotron energy in current experiments at $\nu=5/2$,
(or even larger, see Ref.[\onlinecite{Dean08}])
can, therefore, determine the spin physics of the ground-state
(which is what happens at $\nu=1,1/3$, where the ground-state would be
spontaneously polarized even if the Zeeman energy were precisely zero).
While the Pfaffian and anti-Pfaffian states are fully spin-polarized,
there are also paired states which are not fully-polarized 
\cite{331,Ho95,Dimov,Yang08}, such as the so-called $(3,3,1)$ state.
Therefore, the experiments observing charge $e/4$ quasiparticles do not rule them out.
Experiments which seek to directly probe the
spin polarization at $\nu=5/2$ are inconclusive \cite{Jim}.
Since the proposed non-Abelian
states, whether the Pfaffian or the anti-Pfaffian, are all fully
spin-polarized whereas the competing spin-unpolarized states (e.g. the
hollow-core state or the (331) state) are all Abelian, it becomes imperative
that the issue of 5/2 spin-polarization is resolved by a serious numerical
calculation, which is what we achieve in this work.

\begin{centering}
\begin{figure}
\epsfig {file=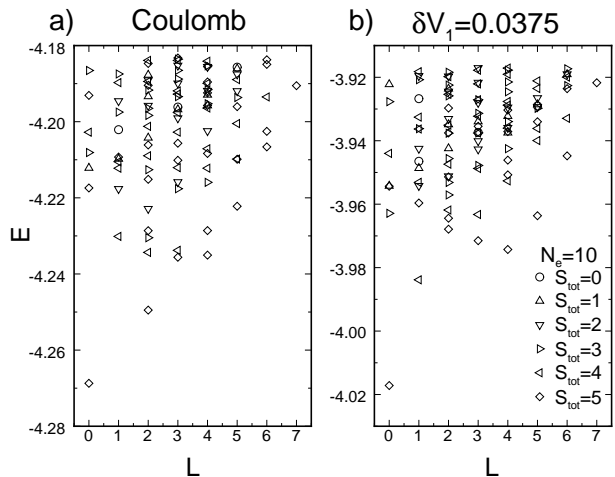,width=65mm,angle=-90}
\caption{Low-energy spectrum of system with $N_e=10$ electrons and shift $S=3$ on the sphere 
obtained with exact diagonalization for: a) Coulomb 
interactions and b) Coulomb interactions with the $V_1$ pseudopotential varied to maximize 
the overlap between the numerical ground-state and 
the Moore-Read state for the case of fully spin-polarized electrons.
}
\label{fig1}
\end{figure}
\end{centering}

For the last 25 years numerical methods have had strong predictive power
in the study of FQH systems, and have become a fundamental validation tool for theories.
In a seminal paper\cite{Morf98}, Morf showed that in a half-filled SLL,
the fully polarized state has lower energy than the spin singlet state in
systems of up to 12 electrons. Based on this result, he argued that the
electrons in the SLL are fully polarized at $\nu=5/2$, which ran counter to the
prevalent view at the time (based on tilted-field experiments \cite{Eisenstein88}).
Later, Park {\it et al} \cite{Park} compared the energies of different
ground-state candidates, and concluded that a polarized Pfaffian is favored
against a polarized composite fermion (CF) sea, and unpolarized composite fermion
sea. Recently, Dimov {\it et al} \cite{Dimov} reached the same conclusion by
comparing the Pfaffian and Halperin's (3,3,1) state \cite{331,Ho95} using variational Monte Carlo.
In all these works, all trial states have energies that are substantially higher than
the unpolarized ground-state energy at $\nu=5/2$ obtained by Morf.

\begin{centering}
\begin{figure}
\epsfig {file=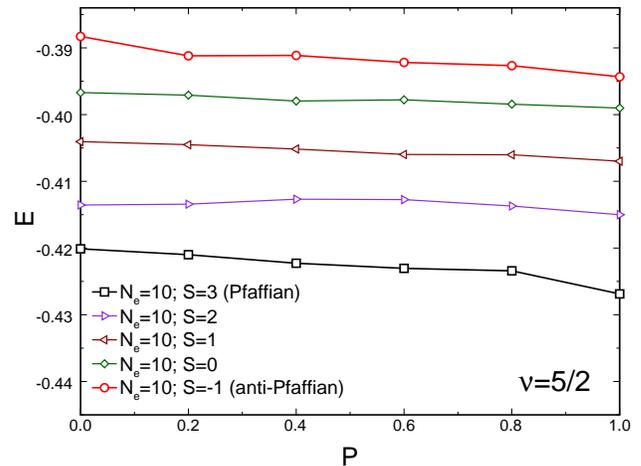,width=65mm,angle=-90}
\caption{
Ground-state energies obtained with DMRG, as a function of polarization $P=2S_{\rm tot}/N_e$, for
$N_e=10$, for the second LL at filling fraction $\nu=5/2$. We show results for a shift $S=3$, corresponding to the Pfaffian, and $S=-1$, 
corresponding to the anti-Pfaffian. We also show results for intermediate shifts. 
Lines are a guide to the eyes.
}
\label{fig2}
\end{figure}
\end{centering}

\section{Method}
The existing numerical evidence suggests that the half-filled SLL is either
fully-polarized, or partially-polarized. However, the latter possibility has
not been explored, probably due to numerical limitations.
In this work we overcome these limitations by combining
exact diagonalization with the recently introduced Density Matrix Renormalization
Group method (DMRG) for studying FQH states on the spherical geometry \cite{Feiguin,Shibata}.
This DMRG approach relies on concepts of exact diagonalization and numerical renormalization 
group, and yields variational results in a reduced 
basis, in the form of a matrix-product state. Contrary to other variational methods, it does 
not rely on an ansatz or prior knowledge of a trial 
wavefunction. The obtained energies are quasi-exact, in the sense that the accuracy is under 
control, and improves as the number of states in 
the basis is increased\cite{dmrg,uli}. We have typically used $4000$ DMRG states, which 
exploits the limits of our computational capability. 

The Hamiltonian that describes a Landau Level is dictated by the Coulomb interaction between 
electrons, making this the quintessential strongly 
correlated problem.
In the spherical geometry, it is written in an angular momentum representation,
which is parametrized by Haldane's
pseudopotentials $V_L$ \cite{Haldane83,Fano} that describe the interaction between
two electrons with relative angular momentum $L$.\cite{foot1}
 In the lowest LL, $V_1$
dominates, explaining why the Laughlin state yields such a good description
at $\nu=1/3$, since it is the exact ground-state of a hard-core Hamiltonian
with $V_L=0$ for $L \neq 1$. However, in the second LL (SLL), the relative
magnitude of the pseudopotentials is such that $V_3$ becomes comparable to
$V_1$, therefore introducing a competition between pairing and Coulomb
repulsion, crucial to stabilize the Pfaffian. (Notice that even-$L$ pseudopotentials only 
become relevant for partially polarized or unpolarized 
states).

%

\section{Results}
Incompressible states at filling fractions $\nu$ are characterized on the
sphere by the number of electrons $N_e$ and flux quanta $N_\Phi$ obeying the
relation $N_\Phi=N_e/\nu-S(\nu)$, where $S(\nu)$ is the so called shift
function. The shift for the Pfaffian $\nu=5/2$ state is $S=3$, and its particle-hole conjugate,
the anti-Pfaffian, is at $S=-1$. In the absence of Landau-level mixing, these states become 
energetically degenerate in the thermodynamic limit.

In Fig. \ref{fig1} we present the low-energy
spectrum of a system with $N_e=10$ electrons obtained 
using exact diagonalization on the sphere at 
half-filling, with the shift $S=3$ corresponding to the Moore-Read (MR) Pfaffian state.  
All values are in units of $e^2/\ell_0$, where $\ell_0=\sqrt{\hbar c/eB}$ is the magnetic length. 
The ground-state is fully magnetized ($S_{\rm tot}=N_e/2=5$), and also has the same orbital 
angular momentum ($L=0$) as the MR state; the 
overlap between the numerical ground-state and the MR state in this case is $70\%$. We also 
find that the full magnetization is a robust 
property of the ground-state when some interaction parameters are varied. In in the same 
figure we present results of the same system with a 
slightly modified Hamiltonian, in which the $V_1$ pseudopotential is tuned to maximize the 
overlap between the numerical ground-state and the 
Moore-Read state for fully spin-polarized electrons; the overlap is $98\%$ in this case. 
(Notice that the overlaps on the sphere are 
larger than on the torus \cite{Rezayi00} and disk \cite{Xin06,Xin08}) Just as in the Coulomb case, the ground-state is 
fully polarized. What is noteworthy about this 
spectrum is that the first excited state has $L=1$ and $S_{\rm tot}=4=N_e/2-1$; this is what 
we expect for the lowest-energy spin-wave 
excitation on top of a fully-magnetized ferromagnetic ground-state. While the spectrum of 
the Coulomb case does not quite show such behavior at 
this particular system size, we believe it is a finite-size artifact; we expect for larger 
system sizes the lowest-energy excitation should be 
a spin-wave, just as we see for $\delta V_1=0.0375$.

\begin{centering}
\begin{figure}
\epsfig {file=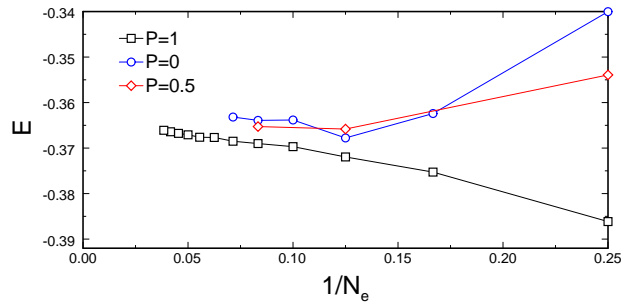,width=42mm,angle=-90}
\caption{
Ground-state energies obtained with DMRG, as a function of $1/N_e$, for
different values of polarization $P$, and shift $S=3$. Energies are in units of the magnetic length and have been rescaled following Ref.[\onlinecite{Morf98}] (see text). Lines are a guide to the eyes.
}
\label{fig3}
\end{figure}
\end{centering}

In Fig.\ref{fig2} we plot the ground-state energies of a system with $N_e=10$ electrons at 
half-filling, as a function of the polarization 
$P=2S_{\rm tot}/N_e$ obtained with the DMRG method. We present results at shift values $S=3$ 
and $S=-1$, corresponding
to the Pfaffian and anti-Pfaffian respectively, and also, for completeness, at intermediate 
values. We have found excellent agreement with exact diagonalization results, with errors in 
the sixth digit, establishing the accuracy of the technique. 
In all cases, the evidence clearly shows that the fully polarized state has lower energy, 
and that the energy increases monotonically with 
decreasing polarization.
For shifts $S=0,1,2$, the energy differences only appear
in the fourth digit. One possible interpretation is that these
values of the shift correspond to excitations above the
Pfaffian and anti-Pfaffian ground-states.
If these excitations were skyrmion-like (i.e. with many 
reversed spins), we would expect the ground-state at these values of the shift to be a spin-singlets.
The addition of a Zeeman energy to the Hamiltonian will even more strongly rule out the
possibility of an unpolarized or even partially-polarized ground-state, even for the lowest 
magnetic field ($\approx$3T) observation \cite{Dean08} of the 5/2 FQH state.

\begin{centering}
\begin{figure}
\epsfig {file=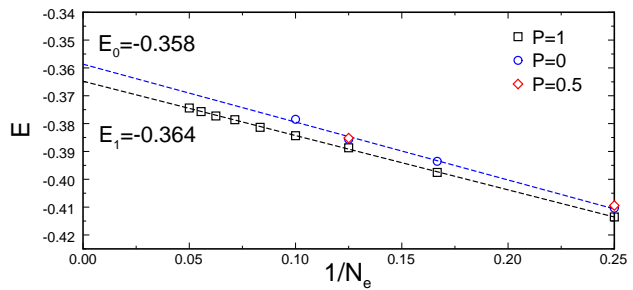,width=42mm,angle=-90}
\caption{
Ground-state energies obtained with DMRG, as a function of $1/N_e$, for
different values of polarization $P$ and shift $S=-1$, corresponding to the anti-Pfaffian. 
Dashed lines indicate a linear extrapolation in $1/N_e$. Energies are in units of the renormalized magnetic length, same as in Fig.\ref{fig3}. (see text)
}
\label{fig4}
\end{figure}
\end{centering}

In Fig.\ref{fig3} we show the ground-state energy as a function of the
number of electrons $N_e$ for different values of the polarization $P$, shift $S=3$, and 
zero Zeeman splitting. 
We have rescaled the energies by a factor $\sqrt{(N_{\Phi}-2)/2N_e}$ to take into account 
finite-size effects on the sphere, 
\cite{Morf98,Morf02}
where we are assuming an 
underlying inert filled ($\nu=2$) lowest Landau level. \cite{foot2}
Our data reproduces the results obtained by
Morf \cite{Morf98} in smaller systems, and we extend the study to $N_e=14$ for the 
unpolarized systems, and $N_e=26$ for the fully polarized states. For
polarization $P=0.5$, we study system sizes up to $N_e=14$.
Notice that the calculations at finite polarization involve a much larger Hilbert space. 
Moreover, the Hamiltonian now includes terms mixing 
spin, making these calculations computationally expensive, and preventing us from reaching 
larger system sizes. Based on extrapolations with the 
number of DMRG states, we estimate our errors to be $10^{-3}$ for the largest systems 
considered, which is of the order of the symbol size.
As previously noticed in Ref.[\onlinecite{Morf98}], the results at finite polarizations exhibit very 
strong finite-size effects. This makes any attempt 
to extrapolate energies to the thermodynamic limit unreliable, even using the larger system 
sizes studied here.

\begin{centering}
\begin{figure}
\epsfig {file=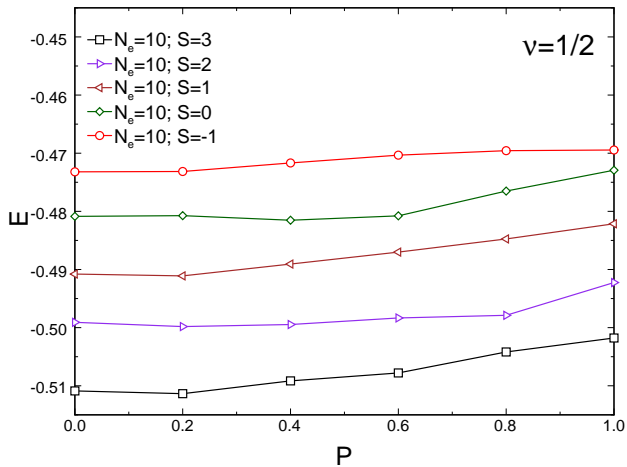,width=65mm,angle=-90}
\caption{
Ground-state energies obtained with DMRG, as a function of polarization $P=2S_{\rm tot}/N_e$, for
$N_e=10$, and different values of the shift.
Results are for the first LL, corresponding to a filling fraction $\nu=1/2$.
Lines are a guide to the eyes.
}
\label{fig5}
\end{figure}
\end{centering}

In Fig.\ref{fig4} we show the ground-state energy as a function of $1/N_e$ for
a shift $S=-1$, corresponding to the anti-Pfaffian.
Notice that this calculation involves four more orbitals than the previous case, making it 
computationally more demanding.
An extrapolation to the thermodynamic limit yields a value of $E(P=1)=-0.364$, identical to 
the best available estimate for the Pfaffian \cite{Feiguin},
as expected for the particle-hole conjugate state. Interestingly,
the partially polarized states show a smoother behavior here than the one observed for 
$S=3$, indicating that finite-size effects may play a less 
important role. This allows one to estimate the ground-state energy of the unpolarized state
in the thermodynamic limit, $E(P=0)=-0.358$. This result is substantially 
lower than the variational energy for the (3,3,1) state, $E_{331}=-0.331$, obtained by Dimov 
{\it et al.} \cite{Dimov}, indicating that the competing unpolarized state may not be a 
a known paired state.  

\begin{centering}
\begin{figure}
\epsfig {file=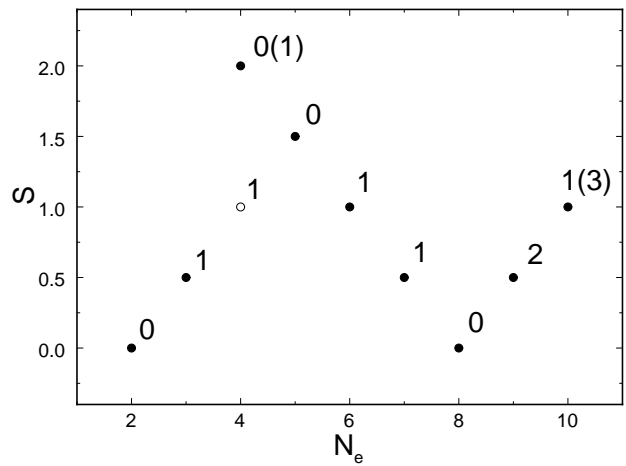,width=65mm,angle=-90}
\caption{
Ground-state spin $S$ for different system sizes, at $\nu=1/2$ and flux $N_\phi=2N_e-2$, corresponding to the composite fermion sea. The numbers next to symbols represent the total angular momentum. Empty symbols correspond to excited states. With the exception of $N_e=4$ and $10$, all the values coincide with those expected from Hund's rule.
}
\label{fig6}
\end{figure}
\end{centering}

\section{Discussion}
In interpreting this data, it is worth remembering that our Hamiltonian
is fully spin-rotation invariant since we do not keep the Zeeman term.
Therefore, any polarization which develops
is a result of spontaneous symmetry breaking and will be accompanied
by gapless Goldstone bosons (i.e. spin waves).
If the ground-state is fully polarized, then the $S_{\rm tot}=N/2$
multiplet will have the lowest energy. The other multiplets will
have energies which are higher by $\sim 1/N$ since the spectrum
of a ferromagnet is $\omega \propto k^2$ as a consequence of the
conservation of the order parameter. If the ground-state
is partially-polarized, then some $0<S_{\rm tot}<N/2$ multiplet will have the
lowest energy. The other multiplets will have energies which are higher by
$\sim 1/N$ since, again, there is a ferromagnetic order parameter which is conserved.
If the ground-state spontaneously breaks spin-rotational symmetry
but does not have a ferromagnetic moment, such as the (3,3,1) state,
then the ground-state in a finite system will be a spin-singlet,
but the gap to other multiplets will be $\sim 1/\sqrt{N}$ since the order
parameter is not conserved. Finally, if the ground-state is a
spin-singlet in the thermodynamic limit,
then the lowest energy state will have $S_{\rm tot}=0$ and there will
be a finite gap to the other multiplets, even in the $N\rightarrow\infty$ limit.
Our data is most consistent with a ferromagnetic ground-state. Extrapolating
to larger system sizes, we expect that the $S_{\rm tot}=N/2$ multiplet will
continue to have the lowest energy, but the gap to other multiplets
will shrink as $\sim 1/N$.

Finally, and for completeness, we calculated the ground-state energies as a function of polarization for a system of $N_e=10$ electrons, at filling fraction $\nu=1/2$, {\it i.e.} in the lowest LL. Results for different shifts are displayed in Fig.\ref{fig5}. The most striking observation is that the ground-state is partially polarized for all the values of shift considered. Thus, the situation at $\nu=1/2$ is very
different from filling fraction $\nu=5/2$, as a result of the difference
between the effective interaction (i.e. the pseudopotentials)
in the lowest and second Landau levels. These results
are in qualitative agreement with calculations of the
Coulomb energies of polarized and unpolarized
trial wavefunctions at half-filling of both
the lowest\cite{Park} and second\cite{Dimov}
Landau levels. Notice that we
have set the Zeeman energy to zero in this calculation.
Since the energy splitting between the partially-polarized
states and the full-polarized state is small for shift $S=2$
(corresponding to the compressible ground
state\cite{Halperin93,Rezayi94}), we expect to be able to tune the system
between partially- and fully-polarized compressible ground-states
at $\nu=1/2$ by increasing the Zeeman energy via a tilted field.
On the other hand, our results lead us to expect that
the plateau at $\nu=5/2$ is fully spin-polarized even for
vanishing Zeeman energy.

Our numerical results for the $\nu=1/2$ state, as shown in
Fig. \ref{fig5}, are completely in agreement with the experimental
findings of Refs. [\onlinecite{Tracy07,Li09}]. Resistive NMR
measurements find that the $\nu=1/2$ plateau is
fully-polarized at high magnetic fields but is not at
low magnetic fields, where it is partially-polarized.
Since the Zeeman energy (relative to the Coulomb energy)
increases with increasing magnetic field, this is
consistent with our numerical findings above. We emphasize
that our prediction for the $\nu=5/2$ state is the opposite
behavior: a fully-spin-polarized ground-state occurs
even for zero Zeeman energy. Therefore, increasing the Zeeman
energy will only make a spin-polarized state more stable
at $\nu=5/2$ and there will not be a Zeeman-energy-induced
transition, in contrast to $\nu=1/2$. The dichotomy between
$\nu=5/2$ and $\nu=1/2$ states is understandable since the
latter is a compressible state while the former is an incompressible
quantized plateau and, therefore, there is no particular reason
for them to have similar spin properties in the ground-state.
Fig. 6 shows the spin $S$ of the ground-state at $\nu=1/2$ and flux $N_\phi=2N_e-2$, plotted vs. system size, corresponding to a composite fermion sea. This pattern is easily seen to follow from Hund’s first rule of maximizing $S$ applied to the angular momentum shells of weakly interacting composite fermions at zero (effective) magnetic field.
  The data is therefore highly suggestive that in the absence of Zeeman gap the $\nu=1/2$ CF state is unpolarized.  The only exception to this rule is at $N_e=4$ where the actual ground-state $S$ is 2 (solid symbol).  However, the difference in energy between this and the Hund’s rule state (open symbol) is 0.000085 (0.004\% of the ground-state energy).  In all likelihood, it is caused by the aliasing of the CF state with the particle-hole conjugate of the Laughlin state for 3 electrons, which is fully polarized, and should be discounted.   Setting aside this case, the second Hund’s rule\cite{Rezayi94} on the angular momentum of the ground-state (listed in Fig. 6 next to the symbols) appears to hold with one, possibly important, exception for $N_e=10$. Here the difference between the actual ground-state at $L=1$ and the Hund's rule state $L=3$ (indicated in parenthesis) is 0.014\% of the ground-state energy and may be more significant.  The pattern of $L$ vs. $N$ should be 01101102332023320\dots, which matches that of Ref.[\onlinecite{Rezayi94}] (if generalized to include spin).   Without further studies, it is difficult to conclude whether this signifies a breakdown of Hund’s second rule or is in fact an isolated exception. Whatever the case, it will not alter the spin polarization of the ground-state.

In conclusion, we have numerically established that the ground-state of the FQH Hamiltonian
at filling fraction $\nu=5/2$, even in the zero Zeeman energy limit, is fully spin-polarized.
We also find, consistent with experimental findings, that the $\nu=1/2$ compressible composite fermion sea state in the lowest Landau level is partially spin-polarized at low magnetic fields, but may become fully polarized at higher magnetic fields due to the Zeeman energy. Thus, $\nu=5/2$ and $\nu=1/2$ states have contrasting spin-polarization properties at low to intermediate magnetic fields.
We believe that our results and the recent findings \cite{Peterson08} of the expected
topological degeneracy on the torus, when taken together with the observation
of charge $e/4$ quasiparticles at $\nu=5/2$ \cite{Dolev08,Radu08}, make a strong case
for the $5/2$ state to be non-Abelian.
Our results should encourage efforts to observe non-Abelian anyons
at this quantum Hall state and use them for topological quantum
computation. 

\acknowledgements
We thank I. Dimov and R. Morf for discussions. The study of $\nu=1/2$ was motivated by conversations with J. Eisenstein.
This research has been supported by Microsoft Station Q,
the NSF under DMR-0704133 (K.Y.) and DMR-0411800 (CN),
the DOE under DE-FG03-02ER-45981 (EHR), and by DARPA-QuEST (SDS and CN).


\begin{thebibliography}{99}
\bibitem{foot1} In order to fix the total spin in 
the DMRG we have added a term $\lambda S^2$ to the 
Hamiltonian, where $\lambda$ is a large constant, and worked with a fixed $S^z_{\rm tot}$.
\bibitem{foot2} In doing Morf's rescaling of the magnetic length
(the so called "ion disk radius") we add the effects of an inert filled LL to the conventional factor $\sqrt{N_\Phi/N_e}$. Either definition is valid, since the rescaling
was meant to make the extrapolation to infinite size look smoother.  In that limit either factor should lead to identical results.
\end{thebibliography}
\end{document}